# Angular Dependence of Spin Torque Critical Currents for CPP-GMR Read Heads

Neil Smith, J. .A. Katine, Jeffrey R. Childress, and Matthew J. Carey,

*Abstract*—This paper employs (analytical) micromagnetic modeling to derive expressions for the critical current at the onset of spin-transfer-torque (STT) instability in CPP-GMR read heads, as a function of the relative angle between the free and reference layer magnetizations, including a general angular dependent STT coefficient. Experimental measurement of the angular dependence of the critical currents are made on 50-nm sized CPP-GMR devices with synthetic antiferromagnet pinned layers, and fabricated using e-beam lithography. The results are consistent with prior theoretical models, but indicate perhaps unanticipated implications for read head operation.

## I. INTRODUCTION

Since first described theoretically [1]-[2], the phenomenon of spin-transfer-torques (STT) from spin-polarized conduction electrons passing current perpendicular-to-plane (CPP) through thin ferromagnetic films has received much recent attention, including theoretical [3]-[5], modeling [6]-[12], and experimental [11]-[18]. This interest in STT has been mostly as a field-*independent* means for magnetization *reversal* in thin magnetic films, particularly that of the free layer in MRAM memory elements. Here, the STT-induced reversal process is between two uniaxial stable states with free and reference layer magnetizations $\hat{m}_{\rm free}$ and $\hat{m}_{\rm ref}$ being collinear. Receiving less attention is the critical current for the onset of undesirable STT-induced noise/oscillations of the free layer in CPP-GMR spin-valve recording heads [19]-[20], where $\hat{m}_{\rm free}$ has *unidirectional* stability and is *orthogonal* to $\hat{m}_{\rm ref}$ in the quiescent state. Receiving relatively little experimental [21] or modeling consideration is the influence here of any intrinsic angular (i.e., $\hat{m}_{\rm free} \cdot \hat{m}_{\rm ref}$) dependence of STT, which was first discussed theoretically by Slonczewski [1]. This paper will first discuss theoretical modeling of the critical currents, and then describe a set of detailed experimental measurements of the *angular dependence* of these critical currents in 50-nm sized CPP-GMR devices.

## II. STABILITY ANALYSIS

The results below are based on the simple model of a CPP spin-valve sensor illustrated in Fig. 1. Each magnetic layer is treated as a uniformly magnetized macro-spin with a unit magnetization vector, $\hat{m} \equiv (m_x, m_y, m_z)$. The magnetization $\hat{m}_{\rm pin}$ of the bottom "pinned" layer is assumed to be exchange pinned along the $+\hat{z}$-axis by an adjacent antiferromagnetic (AF) layer (not shown), and is also strongly AF-coupled to $\hat{m}_{\rm ref}$ of the "reference" layer through a thin Ru spacer. Both $\hat{m}_{\rm pin}$ and $\hat{m}_{\rm ref}$ are assumed rigidly fixed. The free layer magnetization $\hat{m}_{\rm free}(t)$ is treated as a dynamical variable.

In the *absence* of STT effects, the equations of motion for $\hat{m}_{\rm free}(t)$ are taken to be the standard Gilbert equations:

$$d\hat{m}/dt = \gamma(H_{\rm eff} \times \hat{m}) + \alpha\,(\hat{m} \times d\hat{m}/dt)$$
$$\Rightarrow (\hat{m} \times d\hat{m}/dt) + \alpha\,d\hat{m}/dt = \gamma[H_{\rm eff} - (H_{\rm eff} \cdot \hat{m})\hat{m}] \quad (1)$$
$$H_{\rm eff} \equiv -(1/M_s V_{\rm free})\,\partial E_{\rm free}/\partial\hat{m}$$

using $\hat{m} \leftrightarrow \hat{m}_{\rm free}$ interchangeably as clarity permits. $M_s$, $V_{\rm free}$ and $E_{\rm free}(\hat{m})$ are the saturation magnetization, volume and free energy of the free layer, respectively, $\gamma > 0$ is the gyromagnetic ratio, and $\alpha$ is the Gilbert damping parameter.

The physical basis for STT has been described earlier by Slonczewski [1]. From a modeling perspective, the effect of STT on the motion of $\hat{m}_{\rm free}(t)$ was found in [1] to result in an

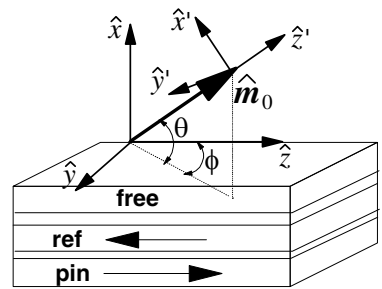

Fig. 1 Cartoon of magnetic layers of CPP-GMR device. *Positive* electron current $I_e > 0$ is defined when *electrons* travel along the $+\hat{x}$ axis. For the rotated $x'y'z'$ coordinate system, the $y'$ axis lies in the $y$–$z$ plane.

added spin-torque contribution to $d\hat{m}/dt$ of the form:

$$(d\hat{m}/dt)_{st} \equiv \beta(\hat{m}_{free} \cdot \hat{m}_{ref}) \gamma H_{st} \hat{m}_{free} \times (\hat{m}_{ref} \times \hat{m}_{free})$$
$$H_{st} \equiv (\gamma \hbar / 2e) \, P I_e /(M_s V_{free}) \quad (2)$$

where $P$ is the polarization of the conduction electrons, and electron current $I_e$ is taken as positive when *electrons* flow from reference layer to free layer (along $+\hat{x}$-axis in Fig. 1). The dimensionless factor $\beta(\hat{m}_{free} \cdot \hat{m}_{ref}) > 0$ is of order unity, with a predicted theoretical form described earlier [1]. It will be treated here in general [10], leaving its angular dependence to be the primary subject of the work of this paper.

Adding the $d\hat{m}/dt$ contribution of (2) to (1), the form of the Gilbert equations *including* STT is unchanged, provided one simply redefines the effective field $H_{eff}$ as:

$$H_{eff} = -(1/M_s V_{free})\partial E_{free}/\partial \hat{m} - \beta(\hat{m}) H_{st} \hat{m}_{ref} \times \hat{m}_{free} \quad (3)$$

The equations of motion can be transformed into a (primed) coordinate system (Fig. 1) where $\hat{z}' \equiv \hat{m}(t=t_0)$ is parallel to $\hat{m}_0 \equiv \hat{m}_{free}(t_0)$ at time $t_0$. The equations hold for $t \geq t_0$ when the *two*-dimensional vector $m' \equiv (m'_x, m'_y)$ is *small*, i.e., $|m'| \perp 1$. The transformation is described by a 3×2 matrix $\vec{R}$ (see Appendix) such that $\hat{m} = \hat{m}_0 + \vec{R} \cdot m'$ to first order in $m'(t)$. The *linearized* equations of motion for $m'(t)$ are

$$\frac{1}{\gamma}\frac{d}{dt}\begin{pmatrix}\alpha & -1 \\ 1 & \alpha\end{pmatrix} \cdot \begin{pmatrix}m'_x \\ m'_y\end{pmatrix} + \vec{H}' \cdot m'(t) = \vec{R}^T \cdot H_{eff}(\hat{m}_0)$$
$$\vec{H}' \equiv \vec{R}^T \cdot \vec{H} \cdot \vec{R}, \; \vec{H} \equiv [(\hat{m} \cdot H_{eff})\vec{1} - (\partial H_{eff}/\partial \hat{m})]\big|_{\hat{m}=\hat{m}_0} \quad (4)$$

where $\vec{H}$ and $\vec{H}'$ are 3×3 and 2×2 Cartesian tensors, $\vec{R}^T$ is the 2×3 transpose of $\vec{R}$, and $\vec{1}$ is the 3×3 identity tensor.

The present interest is in the prediction of the onset of STT-induced dynamical instability at critical current $I_e \to I_e^{crit}$, and in particular, the angular $(\theta, \phi)$ dependence of $I_e^{crit}$ due to that of $\beta(\hat{m}_{free} \cdot \hat{m}_{ref})$. The free-layer reference vector $\hat{m}_0$ is assumed to be an equilibrium for $|I_e| < |I_e^{crit}|$ satisfying $\vec{R}^T \cdot H_{eff}(\hat{m}_0) = 0$. *Stability* of an equilibrium $\hat{m}_0$ requires natural-mode solutions $m' \propto e^{-i\lambda t}$ of (4) with $\text{Im}\,\lambda < 0$, which leads to following two necessary stability conditions [10]:

$$\det \vec{H}' \equiv H'_{xx}H'_{yy} - H'_{xy}H'_{yx} > 0$$
$$\alpha (H'_{xx} + H'_{yy}) + (H'_{yx} - H'_{xy}) > 0 \quad (5)$$

In the absence of STT ($H_{st}=0$), both $\vec{H}$ and $\vec{H}'$ are *symmetric*, and (5) reduces to $\det \vec{H}' > 0$ ($\vec{H}'$ positive-definite). However, when $H_{st} \neq 0$, the $\hat{m}_{ref} \times \hat{m}_{free}$ term in (3) yields a resultant $\vec{H}'$ that is *nonsymmetric* ($H'_{yx} \neq H'_{xy}$). As detailed elsewhere [10], the physical consequences of this nonsymmetry can lead to a systematic transfer of energy from the current $I_e$ to the macro-spin $\hat{m}_{free}$, resulting in an STT-induced instability described by the *second* result of (5).

A simplified model corresponding to the experiment described in Sec. (III) is that of a free-layer with in-plane *unidirectional* anisotropy provided by an in-plane (*y-z* plane in Fig. 1) magnetic field $H_p$, and a uniaxial out-of-plane (shape) anisotropy $H_\perp (\sim 4\pi M_s - H_{k\perp})$. The reference layer $\hat{m}_{ref}$ is also assumed to be in-plane. The derivation steps and linear algebra necessary to evaluate (5) for this case are outlined in the Appendix. The second stability condition of (5) becomes:

$$(2H_p + H_\perp)\alpha + H_{st}[2q\beta(q)-(1-q^2)\beta'(q)] > 0$$
$$H_p \equiv |H_p|, \; q \equiv \hat{m}_0 \cdot \hat{m}_{ref}, \; \beta' \equiv d\beta/dq \quad (6)$$

Unlike the more commonly considered collinear case $|q|=1$, STT-stability is, in general, influenced by both the function $\beta(q)$ and its derivative $d\beta/dq$. This is particularly true in the quasi-orthogonal case $|q = \hat{m}_0 \cdot \hat{m}_{ref}| \ll 1$ that is relevant to the desired nominal bias state of CPP-GMR read heads.

The value of $H_{st}(I_e)$ (see (2)) for which the left side of (6) just vanishes determines the critical current $I_e^{crit}(q)$. By taking the ratio of (6) evaluated at $I_e^{crit}(q)$ and $I_e^{crit}(q=-1)$, the resultant quotient may be arranged in the form:

$$\frac{1-q^2}{2}\frac{d\rho}{dq} - q\rho(q) = F(q) \equiv \frac{(2H_p+H_\perp)|_q}{(2H_p+H_\perp)|_{-1}} \frac{I_e^{crit}(-1)}{I_e^{crit}(q)} \quad (7)$$
$$\rho(q) \equiv \beta(q)/\beta(-1)$$

The result in (7) eliminates physical parameters $\alpha$, $P$, and $V_{free}$, and contains only quantities that are subject to experimental measurement (Sec. IIIB). Since $H_\perp \gg |H_p|$ in practical cases, the dependence of $q(H_p)$ implicit in (7) need not be known to great accuracy. A dual branch solution of the differential equation in (7), chosen to avoid integration across a zero of $F(q)$ (infinity/discontinuity in $I_e^{crit}(q)$) is given by

$$\beta^\pm(q)/\beta(-1) = 2/(1-q^2) \int_{\mp 1}^{q} F(q')\,dq' \quad (8)$$

Using (7) and/or (8) requires measuring $I_e^{crit}$ as a function of $q = \hat{m}_{free} \cdot \hat{m}_{ref}$, which will be discussed further in Sec. III.

## III. Experiment

### A. Film Deposition and Device Fabrication

Samples were fabricated by magnetron sputtering at room temperature on silicon. The base pressure was 2x10-8 Torr. Ar pressure during deposition was 2mT. The film stack consists of (in nanometers): a bottom lead structure of: Ta(5)/Cu(40)/Ta(2), followed by the spin-valve: PtMn(15)/Co$_{84}$Fe$_{16}$(2)/Ru(0.8)/Co$_{84}$Fe$_{16}$(2.2)/Cu(3.5)/Co$_{84}$Fe$_{16}$(1)/Ni$_{88}$Fe$_{12}$(2.4), and finally a top cap of Cu (20)/Ru(4)/Ta (2.5). The thick Cu cap protects the free layer from oxidation during the annealing process, and allows good electrical contact to be made between the top lead and the pillar. The CoFe/Ru/CoFe synthetic-antiferromagnet (SAF) substantially reduces the dipole field on the free layer, facilitating STT measurements with small in-plane fields. The PtMn-SAF exchange coupling keeps the 2.2nm-thick CoFe reference layer well pinned for collinear in-plane fields up to ~2 kOe.

Following film deposition, electron beam lithography was used to pattern a high resolution negative e-beam resist, HSQ. The Fox-12 HSQ formulation (available from Dow Corning) was spun to a thickness of 100 nm. Electron beam exposure essentially converts HSQ into SiO$_2$, whose high resistance to ion milling allows the resist to directly serve as a high-fidelity mask during the etching of our devices. After milling, 100 nm of aluminum oxide was ion-beam deposited onto the wafer, encapsulating the milled pillar and resist. Due to the mechanical instability of the resist mask, the HSQ on top of the pillar is readily removed during a short chemical mechanical polish (CMP), creating a via in the aluminum oxide that allows self-aligned contact to the top of the pillar. To minimize contact resistance, it is important to perform an *in situ* sputter etch prior to depositing the top Cr/Au lead.

### B. Measurement of Critical Currents

Fig. 2 shows $\delta R$ vs. applied field $H_a$ for a 50 nm diameter circular pillar CPP-GMR device ($\approx 35 \Omega$), measured on a Monarch prober. The $\delta R = 0$ point for the $\delta R$–$H_{az}$ reference loop is chosen at $H_{az} = -1 \text{kOe}$ (low-resistance state $\hat{m}_{\text{free}} \cong \hat{m}_{\text{ref}} \cong -\hat{z}$). To compensate for $I_e$-dependent thermal shift, the $\delta R$–$H_{ay}$ curves with $I_e > 0$ are individually zero-shifted to coalesce at large $|H_{ay}|$ with the $I_e = -0.25 \text{mA}$ $\delta R$–$H_{ay}$ loop, which is further aligned at $H_a \cong 0$ with the $\delta R$–$H_{az}$ loop. The latter is *non-hysteretic*, but indicates an internal field $H_i \approx +160\hat{z}$ Oe due to residual magnetostatic coupling between free and ref/pin layers, yielding a high-resistance remnant state ($\hat{m}_{\text{free}} \cong \hat{z}$, $\hat{m}_{\text{ref}} \cong -\hat{z}$) at $H_a = 0$.

Although qualitatively similar with others, this device was chosen for its greater than typical symmetry about $H_{ay} = 0$ in its $\delta R$–$H_{ay}$ loops, as expected from the model of Fig. 1.

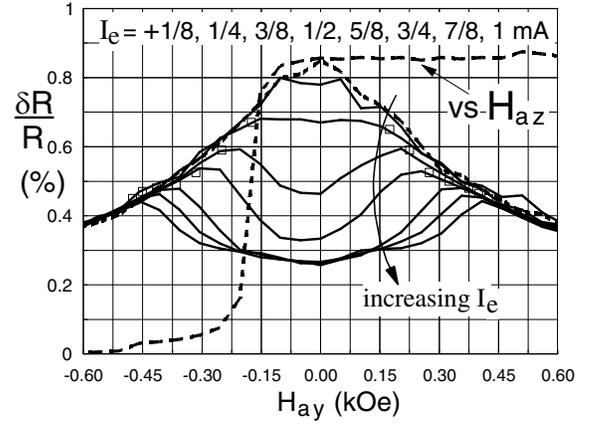

Fig. 2 $\delta R$–$H_{ay}$ loops for various $I_e > 0$ as indicated. The Dashed curves show $\delta R$–$H_{ay}$ and $\delta R$–$H_{az}$ reference loops at $I_e = -0.25 \text{mA}$. The open square symbols indicate estimated critical points.

Noteworthy here is the gross distortion in the $\delta R$–$H_{ay}$ loops, resulting in *minima* at $H_{ay} \approx 0$ which grow broader and deeper with increasing *positive* $I_e$. By contrast, $\delta R$–$H_{ay}$ loops for $-1 \text{mA} \leq I_e$ resemble the $-0.25 \text{mA}$ loop (Fig. 2).

Fig. 3 shows 1-MHz bandwidth rms power spectral density vs. $I_e$, measured at 75 MHz. on a high-frequency prober using an Agilent-E4440A spectrum analyzer. The latter is sweep-triggered by the function generator that drives a 2-Hz sawtooth current into the CPP-GMR device. Subtracting out the $I_e = 0$ electronics noise leaves a residual $\approx 0.15\, (\text{nV}/\sqrt{\text{Hz}})/\text{mA}$ magnetic/thermal background. Of more significance here is the much larger, telegraph-like noise (as revealed by the full spectrum) which onsets sharply for *positive*, $H_{ay}$-dependent "critical" values of $I_e$. Combined, Figs. 2 and 3 strongly suggest the STT-induced critical instability of the unidirectional device modeled in Sec. II. In particular, the *uniquely* stable state $\hat{m}_{\text{ref}} \cong -\hat{z}$, and $\hat{m}_{\text{free}} \approx \hat{H}_p$ becomes

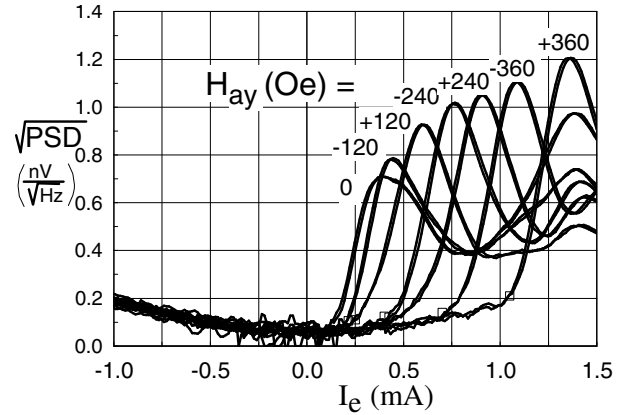

Fig. 3 rms power spectral density (PSD) at 75 MHz vs $I_e$, for various values of $H_{ay}$ as indicated. The open square symbols indicate the estimated critical points. All curves are averaged over ~30 cycles.

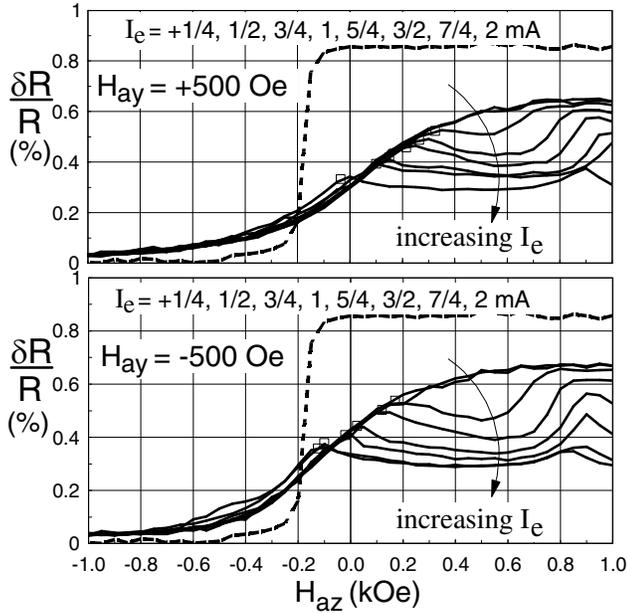

Fig. 4  $\delta R$–$H_{az}$ curves with fixed $H_{ay} = \pm 500$ Oe (as labeled), for various *positive* values of $I_e$. The open square symbols indicate the estimated critical points. The dashed $\delta R$–$H_{az}$ curve is from Fig. 2.

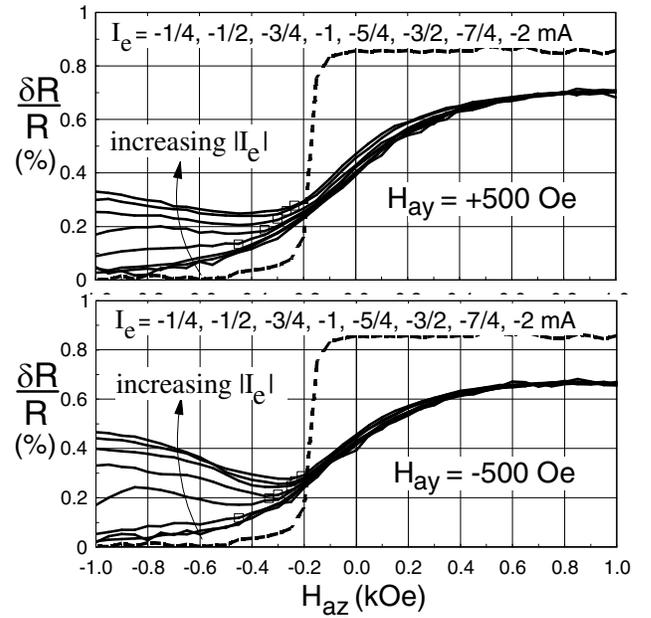

Fig. 5  $\delta R$–$H_{az}$ curves with fixed $H_{ay} = \pm 500$ Oe (as labeled), for various *negative* values of $I_e$. The open square symbols indicate the estimated critical points. The dashed $\delta R$–$H_{az}$ curve is from Fig. 2.

unstable when $I_e > I_e^{\text{crit}} > 0$, after which $\hat{m}_{\text{free}}(t)$ undergoes a continual quasi-chaotic motion indicated by the large-scale noise in Fig. 3. The resultant reduction in the time-averaged value $\langle \hat{m}_{\text{free}} \cdot \hat{z} \rangle \perp 1$ lowers the apparent resistance in $\delta R$–$H_{ay}$ loops of Fig. 2. Comparing Figs. 2-3 shows fairly good agreement in the estimation of the "critical points" $(I_e^{\text{crit}}, H_{ay}^{\text{crit}})$ as determined by the rapid onset of RF noise, or the somewhat less distinctive onset of negative $\delta R$ deviations from the *stable* $\delta R$–$H_{ay}$ loop shape with $I_e = -0.25$ mA. (Critical point evaluation here is somewhat subjective at much beyond ~10% precision.) Figs. 2-3 also indicate some residual $\pm H_{ay}$ asymmetry in $I_e^{\text{crit}}(H_{ay})$ (e.g., $\pm 360$ Oe in Fig. 3).

To simulate the bias conditions for an operating read head, $\delta R$–$H_{az}$ loops are measured with a *fixed* $H_{ay} = \pm 500$ Oe to approach the nominal bias state $\hat{m}_{\text{ref}} \cdot \hat{m}_{\text{free}} \approx 0$, linearizing the response of the device to $z$-axis fields. These loops, for either polarity of $I_e$ are shown in Figs. 4 and 5. Similar to Fig. 2, these $\delta R$ curves are zero-shifted to coalesce at large positive or negative $|H_{az}|$, respectively, and such that the undistorted loops for $|I_e| \leq 0.5$ mA yield the "expected" $\delta R$ value (see Sec. IV) at $H_{az} = -1$ kOe relative to the reference $\delta R$–$H_{az}$ loop. This procedure *independently* achieves fairly consistent values of $\delta R / R \approx 0.4\%$ at $(|H_{ay}| = 500\,\text{Oe}, H_{az} = 0)$ when comparing the four sets of data of Figs. 4-5 with that of Fig. 2.

In Figs. 4-5, the now bipolar (positive in Fig. 4, negative in Fig. 5) critical currents $|I_e^{\text{crit}}(H_a)| > 0.5$ mA can be estimated from the $(I_e, H_a)$ locations of the initial $\delta R$-deviations from the $|I_e| \leq 0.5$ mA loops. As before, these data show some residual $\pm H_{ay}$ asymmetry. Measuring $(I_e^{\text{crit}}, H_a^{\text{crit}})$ here using the PSD required a two-axis field capability that was not available on the high frequency tester at the time of this study.

Fig. 6 illustrates the determination of the critical currents $I_e^{\text{crit}}(q \equiv \hat{m}_{\text{fre}} \cdot \hat{m}_{\text{ref}} = \pm 1)$ PSD measurements. The total $z$-axis field on the free layer is $H_{pz} \cong H_{az} + H_i$, with $H_i \approx +160$ Oe. Though the small difference in apparent $I_e^{\text{crit}}$ by inspection of the $H_{az} = +180$ and $+500$ Oe loops ($H_{pz} = +340, +660$ Oe) is consistent with Eq. (6), the larger difference between the $H_{az} = 0$ and $+180$ Oe loops, and the *gross* difference between the $H_{az} = -500$ and 0 Oe loops ($H_{pz} = -340$ and $-20$ Oe, the latter being magnetothermally metastable) cannot be so accounted for. They are instead due to thermal fluctuations over ~0.1 sec measurement times [18]. As these data suggest thermal effects are small for $|H_p| \geq 300$ Oe, the values $I_e^{crit}(q=-1) \cong +0.2$ mA, and $I_e^{crit}(q=+1) \cong -0.6$ mA (at $|H_p| \cong 340$ Oe) will be used for the analysis in Sec. IV.

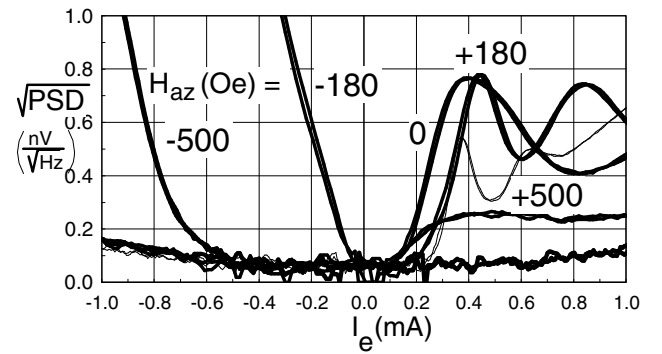

Fig. 6 PSD at 75 MHz vs $I_e$ at $H_{ay} = 0$, $H_{az}$ as indicated. The fine-lined curve is for $H_{az} = +500$ Oe. All curves are averaged over ~30 cycles.

## IV. ANALYSIS AND DISCUSSION

Both original [1] and later [5] theoretical formulations of STT by Slonczewski predict the following form [22] for $\beta(q)$:

$$\beta(q)/\beta(-1) = (1-A)/[1+Aq] \quad (9)$$

For the present CPP-GMR film stack, the estimate $A \cong 0.5$ for the constant $A$ follows from the measurements of Fig. 6, and the relationship $\beta(+1)/\beta(-1) = I_e^{crit}(-1)/I_e^{crit}(+1) \cong 1/3$ from (7). However, theory ([5], [23]) also predicts a direct relation between the q-dependence of β (i.e., $A \neq 0$) and a *nonlinear* q-dependence for the CPP-GMR resistance variation:

$$\frac{\delta R(q)}{R} = \frac{\Delta R}{R} \frac{(1-q)/2}{1+[A/(1-A)](1+q)} \quad (10)$$

taking $\delta R(q=+1) \equiv 0$, as was specifically done by zero-shifting the measured $\delta R$–$H_a$ loops as described in Sec. III.B. Similar angular dependence of CPP-GMR was also recently measured in magnetic multilayers at 4.2 K [24].

Equation (10) may be exploited to determine $q = q_{crit}$ by using the set of ~45 triplets of $(I_e^{crit}, H_p^{crit}, (\delta R/R)_{crit})$ from the measurements of Sec. III.B, substituting $(\delta R/R)_{crit}$ into (10) (with $\Delta R/R \cong 0.87\%$), and inverting to solve for $q_{crit}$. This yields ~45 discrete values for $I_e^{crit}(q)$, from which one obtains $F(q)$ using the right side of (7) (with $H_\perp \approx 7.5$ kOe [18]). The results of this procedure are plotted in Fig. 7.

Also included in Fig. 7 are results with $q_{crit}$ evaluated as

$$\hat{m}_{ref} = -\hat{z} \rightarrow q_{crit} \cong -\hat{z} \cdot \hat{H}_p^{crit} \quad (11)$$

using (A2). However, (11) assumes circular pillars with negligible in-plane anisotropy, zero interlayer coupling, and complete spatial uniformity of $\hat{m}_{free}$, none of which is assumed for (10). The use of (11) is expected to be most valid for critical points with the largest $|H_a|$, such as the larger $|q_{crit}|$ data extracted from Figs. 4,5. Fig.7 confirms that the results from (10) and (11) agree best for these data points. Using (10), the complete set of critical points (from $\delta R$–$H$ or PSD) in Fig. 7 appear to lie on a single curve.

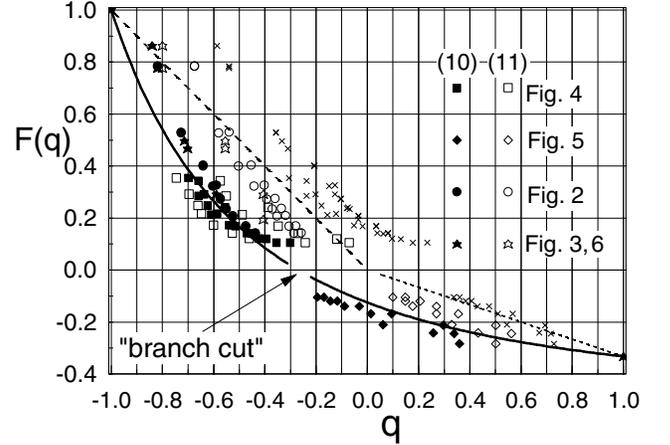

Fig.7. $F(q)$ using the data from Figs. 2-6 (as indicated), evaluated from the *right* side of (7), with $q$ from (10) (filled symbols) or (11) (open symbols). The smooth curves are generated from the *left* side of (7), with $\beta(q)$ from (9) (solid) or (12) (dashed). The points denoted with "×" are generated by evaluating $q$ using (10) with $A \equiv 0$.

Most noteworthy in Fig. 7 is the consistent agreement between the measured and predicted $q = \hat{m}_0 \cdot \hat{m}_{ref}$ dependence when using (7) with $\beta(q)$ from (9), *and* using (10) (with $A = 0.5$) to measure $q_{crit}$. (Deviations near $q \approx -0.8$ likely result from thermally reduced $I_e^{crit}$ for these lowest $|H_p|$ critical points from Figs. 2,3.) In particular, the predicted location of the "branch-cut" (see (8)) at $q \cong -0.27$ agrees remarkably well with the apparent absence of measurable critical points at/near the zero of $F(q)$. Its negative shift from $q = 0$ results from the $d\beta/dq$ term first predicted in (6).

The dashed curve in Fig. 7 was computed from (7) using the following simple but *discontinuous* functional form for $\beta(q)$:

$$\beta^\pm(q)/\beta(-1) = 1/3 - (2/3)\text{sgn}(q) \quad (12)$$

chosen to provide a rough approximation to the "×" critical points that were extracted with the *linear* form of (10) (with $A \equiv 0$) universally used for *CIP*-GMR devices. The nonphysical discontinuity of this extracted $\beta(q)$ is further indicative of the physical connection between STT and MR

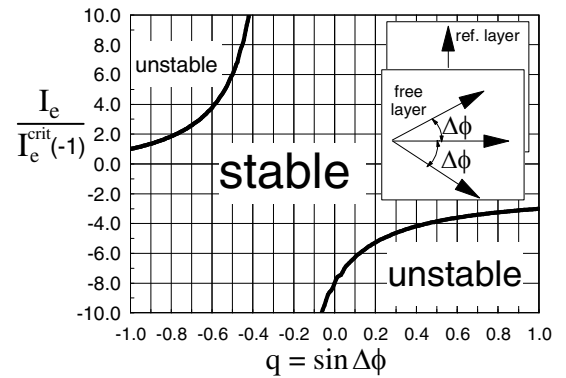

Fig. 8. $q$–$I_e$ stability space for the CPP-GMR read head configuration illustrated in the upper right, and computed as described in the text.

in *CPP* devices, as is implied by (9) *and* (10).

An important practical finding of this study is the heretofore unconsidered role of $d\beta/dq$ in governing the STT-induced instability in CPP-GMR read heads, as described here in (6). It can lead to some non-intuitive consequences regarding stable operational ranges of such devices. This is illustrated in Fig. 8, which delineates stability regions in $q$–$I_e$ space. The boundary curves are determined by solving (7) for $I_e^{\text{crit}}(q)$ in terms of the differential functional of $\beta(q)$, and then using $\beta(q)$ from (9) with $A = 1/2$. (For simplicity, it is assumed here that $|H_p^{\text{crit}}|/H_\perp \ll 1$ and/or is slowly varying with $q$.)

For example, assuming a moderate head excitation $|q = \sin\Delta\phi| \leq 0.4$ about the bias point $\phi = 0$, stable operation with STT present is predicted for bias currents satisfying $-4 \leq I_e^{\text{bias}}/I_e^{\text{crit}}(-1) \leq +10$. Hence, for stable operation with maximum $I_e^{\text{bias}}$ (and thus signal), a substantial 2.5-fold advantage is achieved by using *positive* $I_e^{\text{bias}}$, despite (since STT $\propto \beta$) the fact that $\beta(q)$ *increases* monotonically with *decreasing q*, and that $q_{\text{crit}}(I_e > 0) < q_{\text{crit}}(I_e < 0)$. Due to the nonlinear $q$-dependence in (10), signal symmetry considerations may lead to other bias point choices.

## APPENDIX

To *first* order in $\boldsymbol{m}' \equiv (m_x', m_y')$, $\hat{\boldsymbol{m}} = \vec{\Re} \cdot \hat{\boldsymbol{m}}'$, with $m_z' = 1$, and 3×3 partitioned rotation matrix $\vec{\Re} = (\vec{R} \mid \hat{\boldsymbol{m}}_0)$ [25]:

$$\hat{\boldsymbol{m}}_0 = \begin{pmatrix} \sin\theta \\ \cos\theta\sin\phi \\ \cos\theta\cos\phi \end{pmatrix}, \vec{R} = \begin{pmatrix} \cos\theta & 0 \\ -\sin\theta\sin\phi & \cos\phi \\ -\sin\theta\cos\phi & -\sin\phi \end{pmatrix} \quad (A1)$$

Left-multiplying (1) by $\vec{\Re}^{-1} \equiv \vec{\Re}^{\mathsf{T}}$ yields, to 1st order in $\boldsymbol{m}'$:
$(\hat{z}' \times d\hat{\boldsymbol{m}}'/dt) + \alpha \, d\hat{\boldsymbol{m}}'/dt = \gamma \vec{\Re}^{\mathsf{T}} \cdot [\boldsymbol{H}_{\text{eff}} - (\boldsymbol{H}_{\text{eff}} \cdot \hat{\boldsymbol{m}})_0 \vec{\Re} \cdot \hat{\boldsymbol{m}}']$.
The $z'$ component of this equation is null. Expressing $\boldsymbol{H}(\hat{\boldsymbol{m}}) = \boldsymbol{H}(\hat{\boldsymbol{m}}_0) + \partial\boldsymbol{H}/\partial\hat{\boldsymbol{m}} \cdot (\hat{\boldsymbol{m}} - \hat{\boldsymbol{m}}_0 \equiv \vec{R} \cdot \boldsymbol{m}')$ then yields (4).

For a free-layer with *out-of-plane* uniaxial anisotropy $H_\perp$, and a unidirectional *in-plane* field $\boldsymbol{H}_p = (0, H_{py}, H_{pz})$, $E_{\text{free}}/M_s V_{\text{free}} = \tfrac{1}{2} H_\perp m_x^2 - H_{py} m_y - H_{pz} m_z$. Assuming an *in-plane* $\hat{\boldsymbol{m}}_{\text{ref}} = (0, \sin\psi, \cos\psi)$, one can, using (3), (A1), and some matrix algebra, evaluate all quantities needed in (4).

The equilibrium $\vec{R}^{\mathsf{T}} \cdot \boldsymbol{H}_{\text{eff}}(\hat{\boldsymbol{m}}_0) = 0$ approximately satisfies:

$$\begin{aligned}\phi &\approx \tan^{-1}(H_{py}/H_{pz}) \\ \theta &\approx \beta_0 H_{\text{st}} \sin(\phi - \psi)/(|\boldsymbol{H}_p| + H_\perp)\end{aligned} \quad (A2)$$

where $\beta_0 \equiv \beta(\hat{\boldsymbol{m}}_0 \cdot \hat{\boldsymbol{m}}_{\text{ref}})$. In practical circumstances, $\theta \ll 1$. Approximating $\hat{\boldsymbol{m}}_0 \cdot \boldsymbol{H}_p \cong |\boldsymbol{H}_p|$ from (A2), the $\vec{H}'$ tensor can be found using (3), (4), (A1), and some matrix algebra. The relevant quantities required for (5) are found to be:

$$\begin{aligned} H'_{xx} + H'_{yy} &= 2|\boldsymbol{H}_p| + (1 - 3\sin^2\theta)H_\perp \\ H'_{yx} - H'_{xy} &= [2q\beta_0 - (1-q^2)d\beta_0/dq]H_{\text{st}}\end{aligned} \quad (A3)$$

where $q \equiv \hat{\boldsymbol{m}}_0 \cdot \hat{\boldsymbol{m}}_{\text{ref}} = \cos\theta\cos(\phi - \psi)$. The $d\beta/dq$ term in (A3) originates from the $\hat{\boldsymbol{m}}$-dependence of $\beta(\hat{\boldsymbol{m}} \cdot \hat{\boldsymbol{m}}_{\text{ref}})$, and its contribution from (3) to $\partial \boldsymbol{H}_{\text{eff}}/\partial\hat{\boldsymbol{m}}$ in (4). Neglecting terms of order $\theta^2$ (see (A2)) then yields the result in (6).